\documentstyle[preprint,aps,tighten,eqsecnum]{revtex}
        \def\be{\nopagebreak[3]\begin{equation}}
        \def\ee{\end{equation}}
        \def\ba{\nopagebreak[3]\begin{eqnarray}}
        \def\ea{\end{eqnarray}}
        \def\nl{\nonumber \\}
        
        \def\d{{\rm d}}
        \def\w{\Omega}
        \def\G{\Gamma}
        \def\Lt{{\cal L}_t}
        \def\tP{\tilde{\Pi}}
        \def\x{\bar{x}}
        \def\H{{\cal H}}
        \def\F{{\cal F}}
        \def\S{{\cal S}}
        \def\a{\alpha}
        \def\c{{\cal C}}
\newcommand{\T}{\textstyle}
\newcommand{\R}{{\sf R\hspace*{-0.9ex}\rule{0.15ex}%
       {1.5ex}\hspace*{0.9ex}}}

\newcommand{\C}{{\sf C\hspace*{-0.9ex}\rule{0.15ex}%
       {1.3ex}\hspace*{0.9ex}}}
\newcommand{\teta}{\rlap{\lower2ex\hbox{$\,\tilde{}$}}\eta{}}
\newcommand{\tieta}{\tilde{\eta}}
       
\begin{document}
\draft
\title{Introduction to the Fock Quantization of the Maxwell Field}
\author {Alejandro Corichi\thanks{corichi@nuclecu.unam.mx}}
\address{Instituto de Ciencias Nucleares\\
Universidad Nacional Aut\'onoma de M\'exico\\
A. Postal 70-543, M\'exico D.F. 04510, MEXICO \\
}
\maketitle
\begin{abstract}
In this article we give an
introduction to  the Fock quantization
of the Maxwell field.
At the classical level, we treat the
theory in both the covariant and canonical phase space formalisms.
The approach is general since we consider arbitrary (globally-hyperbolic)
space-times. The Fock quantization is shown to be equivalent to the
definition of a complex structure on the classical
phase space. As examples, we consider stationary space-times as well as
ordinary Minkowski space-time.
The account is pedagogical
in spirit and is tailored to beginning graduate students. The paper is self
contained and is intended to fill an existing gap in the literature.

\end{abstract}

\section{introduction}


The motivation to write this article comes from the author's
discomfort with the usual treatment that textbooks give to the
{\it canonical quantization} of free fields in their first chapters
\cite{textbook}. There seems to be a ``quantum jump" from the
quantization of mechanical systems with a finite number of degrees of
freedom to the quantization of fields. Here, by fields we mean that
the classical system to be quantized is described by (at least) one
function of space-time. The best known example is precisely the
electro-magnetic field, described by six quantities at each space-time
point.
In  ordinary quantum mechanics,
one starts with the phase space $\G$ of the system, which is normally
given by pairs $(q^i,p_i), \; i=1,2,\ldots,n$ of generalized coordinates
and their conjugate momenta.
The quantization procedure implies a passage from the basic
Poisson Brackets (PB) relations $\{q^i,p_j\}=\delta_j^i$ to the Canonical
Commutation Relations (CCR): $[\hat{q}^i,\hat{p}_j]=i\hbar\delta^i_j$.
This is usually called the {\it Dirac quantization condition}. 
One finally finds a Hilbert space $\H$ and a representation of
the basic observables $\hat{q}^i$ and $\hat{p}_i$ as self-adjoint
operators on $\H$
satisfying the CCR. More precisely, one should find at the classical level
a set $\S$ of {\it elementary observables} (real
functions) on $\G$ that are: i) large enough to generate, via linear
combinations of products of them, any function on $\G$ and ; 2) small
enough to be closed under Poisson Brackets \cite{tate}. To these
observables, elements of $\S$,
there will be associated a quantum operator in a unique way,
satisfying the Dirac quantization condition.
For details see Sec.~\ref{sec0}.

When the classical system to be quantized is a field theory, one is led
to ask: Can we follow the same prescription? that is, can we identify
the phase space of the problem and a set $\S$ of basic observables?
How is the Poisson bracket defined? Can we implement the Dirac quantization
condition and find representations of the CCR? If yes, which is the
Hilbert Space $\H$? The aim of this paper is to give answers to all this
questions when the classical system to be quantized is the free Maxwell
field. In the case of a Klein-Gordon field, the problem is
satisfactorily addressed by Wald \cite{wald2} (The reader is
urged to read the first three chapters of that book). Recall that the Klein
Gordon field is described by a scalar field $\Phi$ on space-time satisfying
the Klein-Gordon equation: $(\Box^2  -m^2)\Phi=0$.
The main difference
between the Klein-Gordon and the Maxwell field is gauge invariance.
This in turn brings some subtleties to the program of quantization. These
problem are dealt with in this paper.

The particular quantization method we shall consider is the one known as Fock
quantization. The intuitive idea is that the Hilbert space of the theory
is constructed from ``n-particle states". (In certain cases one is
justified to interpret the quantum states as consisting of n-particle
states. For a discussion see below.)
As we shall see later, the Fock quantization is naturally constructed
from {\it solutions to the classical equations of motion} and
relies heavily on the linear structure of the space of solutions
(The Klein-Gordon and Maxwell equations are linear).
Thus, it can only be implemented for quantizing {\it linear (free)
 field theories}. The main steps of the quantization are the following: 
Given a 4-dimensional globally hyperbolic space-time $(M,g)$\footnote{
Recall that a globally hyperbolic spacetime is one in which the entire
history of the universe can be predicted from conditions at the instant
of time represented by a hyper-surface $\Sigma$. In technical terms
$\Sigma$ is a Cauchy surface. For details see \cite{wald1}.},
the first step is to consider the vector space $\bar{\G}$ of solutions
of the equations of motion and construct from it the vector space of
{\it physically indistinguishable} states $\G$.
 One then constructs the algebra  $\S$ of
fundamental observables to be quantized, which in this case consists
of suitable linear functionals on $\Gamma$. The next step is to
construct the so called {\it one-particle Hilbert space} ${\cal H}_0$
from the space ${\Gamma}$. As mentioned before, the one particle Hilbert
space ${\cal H}_0$ receives this name since it can be interpreted as
the Hilbert space of a one particle relativistic system
(in the electro-magnetic case, the photon).
The one-particle space is constructed by defining a complex structure
on $\Gamma$ compatible with the naturally defined symplectic structure
thereon, in order to define a Hermitian inner product on $\Gamma$. The
completion with respect to this inner product will be the one-particle
Hilbert space ${\cal H}_0$.
 From the Hilbert space ${\cal H}_0$ one constructs its symmetric
(since we are considering Bose fields) Fock space ${\cal F}_{s}({\cal H})$,
the Hilbert space of the theory.
The final step is to represent the algebra $\S$ of observables in the Fock
space as suitable combinations of
(naturally defined) creation and annihilation operators.
   
We will construct in detail the quantization outlined above for the
case of the Maxwell field. In our opinion, an unified treatment
(although completely elementary) is not available elsewhere.
The structure of the paper is as follows. In Sec.~\ref{sec0} we
give an overview of the prerequisites to tackle the quantization
program. In particular, we review the canonical quantization
using symplectic language.
 In the Sec.~\ref{C.1} we
consider the classical treatment of the Maxwell field. We follow two
paths in the phase space description of the theory. The first one,
the so called {\it covariant phase space} starts from the solutions
to the equations of motion. The second approach, the `standard' $3+1$
formulation, is considered next and compared to the covariant framework.
Sec.~\ref{C.2} addresses the quantization. We outline the quantization
strategy starting from the classical analysis and show that it depends
on certain extra structure (a complex structure) defined on the classical
phase space. We consider then two examples of particular interest on
Minkowski space-time: the standard `positive frequency' decomposition
and the self dual decomposition.
We end with a discussion in Sec.~\ref{C.3}.

Throughout the paper, we use Penrose's  abstract index notation\footnote{In
this notation, the index `$a$' of a vector $v^a$ is to be seen as a
label indicating that $v$ is a vector (very much like the arrow in
$\vec{v}$), and it does not take values in any set. That is,
`$a$' is  not the component of $\vec{v}$ on any basis. For
details see \cite{penr,penr2,wald1}}, and units in which $c=1$, but  keep
$\hbar$ explicit.

\section{Preliminaries}
\label{sec0}

In this section we shall present some background material, both in
classical and quantum mechanics. This section has two parts. In the
first one 
we will introduce some basic notions of symplectic geometry
that play a fundamental role in the Hamiltonian description of
classical systems. In the second part we outline the canonical
quantization starting from a classical system as described in
Sec.~\ref{sec0.1}.

\subsection{Classical Mechanics}
\label{sec0.1}

A physical system is normally represented, at the classical level, by a {\it
phase space}. This consists of a manifold  $\Gamma$ of
dimension $dim (\Gamma )=2n$. Physical states are represented by
the points on the manifold. Observables are smooth, real valued functions
on $\Gamma$. There is a non-degenerate, closed  {\it symplectic}
two-form $\Omega$ defined on it.
 The two-form $\Omega _{ab}$
satisfies: $\nabla_{[c}\Omega _{ab]}=0$, and if $\Omega _{ab}V^b=0$ then
$V^b=0$. Therefore, there exists an inverse $\Omega ^{ab}$ and it defines
an isomorphism between the cotangent and the tangent space at each point of 
$\Gamma$. Here square brackets over a set
of indices means antisymetrization. That is $A_{[ab]}:=\frac{1}{2}(A_{ab}
-A_{ba})$ (and $A_{(ab)}:=\frac{1}{2}(A_{ab}+A_{ba})$). The space $\G$ with
the symplectic two-form $\w$ is called a Symplectic space and denoted by
$(\G,\w )$.

A vector field $V^a$ generates infinitesimal canonical transformations if it
Lie drags the symplectic form, i.e.:
\begin{equation}
{\cal L}_V\Omega =0
\end{equation}
This condition is equivalent to saying that locally the symplectic form
satisfies:
$V^b=\Omega^{ba}\nabla_a f:= X^b_f$, for some function $f$. The vector
$X^a_f$ is called the {\it Hamiltonian vector field of $f$
(w.r.t. $\Omega$)}.
Note that the symplectic structure gives us a mapping between functions on
$\Gamma$ and Hamiltonian vector fields. 
Thus, functions on phase space (i.e. observables)
are generators of infinitesimal canonical transformations.

The Lie Algebra of vector fields induces a Lie Algebra structure on the
space of functions.
\begin{equation}
\{ f,g\} :=\Omega_{ab} X^a_f X^b_g = \Omega^{ab}\nabla_af\nabla_bg
\end{equation}
such that
$X^a_{\{ f,g\} }=-[X_f,X_g]^a$. The `product' $\{\cdot ,\cdot\}$ is
called {\it Poisson Bracket} (PB).

Note that the Poisson bracket $\{f,g\}$ gives the change of $f$ given 
by the motion generated by (the HVF of) $g$, i.e,
\be
\{ f,g\}={\cal L}_{X_g}f
\ee
The PB is antisymmetric so it is also (minus) the change of $g$ generated
by $f$.

The role of the symplectic structure $\Omega$ in symplectic geometry
is somewhat similar to the role of the metric in Riemannian geometry.
It provides a one to one mapping between vectors and one-forms at
each point of the manifold.
There is however a very important difference: In symplectic geometry
one can always find coordinates $(q^i,p_j)$ in
a finite neighborhood such that the symplectic form takes the
canonical form (known as Darboux Theorem),
\be
\Omega_{ab}=2\nabla_{[a}p_i\nabla_{b]}q^i
\ee
With this form, the Poisson bracket between the coordinate functions takes
the form,
\ba
\{q^i,p_j\}=\Omega^{ab}\nabla_a(q^i)\nabla_b(p_j) &=& \delta^i_j\\
\{q^i,q^j\}=\Omega^{ab}\nabla_a(q^i)\nabla_b(q^j) &=&
\{p_i,p_j\}=\Omega^{ab}\nabla_a(p_i)
\nabla_b(p_j)=0
\ea
In such a chart, the $q^i$ coordinates  are like `position' and
$p_i$ are like `momenta'.

Since the symplectic form is closed,  it can be obtained locally from a
{\it symplectic potential} $\omega_a$,
\begin{equation}
\Omega_{ab}=2\nabla_{[a}\omega_{b]}
\end{equation}

Time evolution is given by a vector field $h^a$ whose integral curves 
are the dynamical trajectories of the system. On phase space there is a
{\it preferred} function, the {\it Hamiltonian} $H$ whose Hamiltonian vector
field corresponds precisely with $h^a$, i.e.,
\begin{equation}
h^a=\Omega ^{ab}\nabla_b H \label{hameq}
\end{equation}

Adopting the viewpoint that all observables generate canonical
transformations
we see that the motion generated by the Hamiltonian corresponds to `time
evolution'. The `change' in time of the observables will be simply given
by the Poisson bracket of the observable with $H$:
$\dot{g}:=h^a\nabla_a g=\Omega^{ac}\nabla_c H\nabla_a g=\{ g,H\}$.

If the system has a configuration space $\c$, then the phase space $\G$ is
automatically ``chosen'' to be the cotangent bundle of the configuration
space $T^*\c$. There is also a preferred  1-form on $\c$ that can be
lifted to $T^*\c$  and
taken to be the symplectic potential which determines uniquely the symplectic
structure. Therefore, the fact that there exists a configuration space picks 
for us the phase space and the symplectic two-form. For field theories
this description is obtained when one performs a $3+1$ decomposition on
space-time and the phase space is defined from the initial data of
the theory. An alternative is to consider the covariant variational
principle, without any decomposition,
and construct a naturally defined symplectic
two-form. This is the covariant phase space formalism that will be seen
in Sec.~\ref{C.1}. 

Let us look in detail at the simplest example: a particle in 3 dimensional
Euclidean space. The state of the system is specified by the value of
its configuration $q^i$ and its momenta variables $p_i$. In this case,
$q^i$ are coordinates in the configuration space $\c$. Here $i=1,2,3$
and the dimension of $\Gamma$ is $6$. The phase space has in the case a
cotangent bundle structure $\Gamma=T^*\c$, and the naturally defined
symplectic potential is,
\be
\omega_a=p_i\nabla_aq^i
\ee
from which the natural symplectic structure can be derived,
\be
\Omega_{ab}=2\nabla_{[a}p_i\nabla_{b]}q^i\label{2-form}
\ee
That is, in the dual basis $\{\nabla_a q^i,\nabla_ap_i\}$ for the cotangent
space the 2-form (\ref{2-form}) has a matrix representation that
can be written as,
\[
\Omega_{\underline{a}\underline{b}}=\left(\begin{array}{cc}
                                           0&-I_{n\times n}\\
                                           I_{n\times n}&0 \end{array}
                                    \right)
\]
In the basis of the tangent space to $\Gamma$, the inverse of the
symplectic two-form is given by,
\be
\Omega^{ab}=2\left(\frac{\partial}{\partial q^j}\right)^{[a}\left(
\frac{\partial}{\partial p_j}\right)^{b]}
\ee
The Poisson bracket in this coordinates has the usual form,
\be
\{ f, g\}=\frac{\partial f}{\partial q^i}\cdot
\frac{\partial g}{\partial p_i}-\frac{\partial g}{\partial q^i}\cdot
\frac{\partial f}{\partial p_i}  
\ee
and the evolution equations are
\be
\dot{q}^i=\{q^i,H\}=\frac{\partial H}{\partial p_i}\qquad\mbox{and}
\qquad \dot{p}_i=\{p_i, H\}=-\frac{\partial H}{\partial q^i}
\ee
In this form, we recover the usual textbook treatment of Hamiltonian
mechanics.

In the case that the system exhibits come gauge freedom in the classical
theory, its description in symplectic language gets modified. The details
are different for the covariant and canonical phase space descriptions, but
the common theme is that the phase space accessible to the system is not
a true symplectic space: the two-form $\w$ is degenerate. In this case
the space is called a pre-symplectic space and $\w$ is a called a
pre-symplectic structure. In Sec.~\ref{C.1} we treat the Maxwell system
and comment on the strategy to deal with gauge systems in both
descriptions. Let us now look at the quantization.

\subsection{Quantization}

In very broad terms, by quantization one means the passage from a classical
system, as described in the last part, to a quantum system. Observables
on $\Gamma$ are to be promoted to self-adjoint operators on a Hilbert
Space. However, we know that not all observables can be promoted
unambiguously to quantum operators satisfying the CCR. A well known
example of such problem is factor ordering. What we {\it can} do is to
construct a subset ${\cal S}$ of {\it elementary classical variables}
for which the quantization
process has no ambiguity. This set ${\cal S}$ should satisfy two properties:
\begin{itemize}
\item ${\cal S}$ should be a vector space large enough so that every
(regular) function on $\Gamma$ can be obtained by (possibly a limit of) sums
of products of elements in
${\cal S}$. The purpose of this condition is that we want that enough
observables are to be unambiguously quantized.

\item The set ${\cal S}$ should be small enough such that it is closed
under Poisson brackets. 

\end{itemize}

The next step is to construct an (abstract) quantum algebra ${\cal A}$ of
observables from the vector space ${\cal S}$ as the free associative
algebra generated by ${\cal S}$ (for a definition and discussion of
free associative algebras see \cite{geroch2}). It is in this quantum
algebra ${\cal A}$ that we impose the Dirac quantization condition:
Given $A,B$ and $\{A,B\}$ in ${\cal S}$ we impose,
\be
[\hat{A},\hat{B}]=i\hbar\widehat{\{A,B\}}\label{diracc}
\ee
It is important to note that there is no factor order ambiguity in
the Dirac condition since $A,B$ and $\{A,B\}$ are contained in ${\cal S}$
and they have associated a unique element of ${\cal A}$.

The last step is to find a Hilbert space $\H$ and a representation of the
elements of ${\cal A}$ as operators on $\H$. For details of this approach
to quantization see \cite{tate}.

In the case that the phase space $\Gamma$ is a linear space,
there is a particular simple
choice for the set $\S$. We can take a global chart on $\Gamma$ and
we can choose $\S$ to be the vector space generated by {\it linear} functions
on $\Gamma$. In some sense this is the smallest choice of
$\S$ one can take. As a concrete case, let us look at the example of
$\c=\R^3$. We can take a global chart on $\Gamma$ given by $(q^i,p_i)$ and
consider $\S=\mbox{Span}\{1,q^1,q^2,q^3,p_1,p_2,p_3\}$. It is a seven
dimensional vector space. Notice that we have included the constant
functions on $\Gamma$, generated by the unit function since we know
that $\{q^1,p_1\}=1$, and we want $\S$ to be closed under PB.

We can now look at linear functions on $\Gamma$. Denote by $Y^a$ an
element of $\Gamma$, and using the fact that it is linear space, $Y^a$
also represents a vector in $T\Gamma$. Given a one form $\lambda_a$, we can
define a linear function of $\Gamma$ as follows:
$F_\lambda(Y):=\lambda_aY^a$. Note
that $\lambda$ is a label of the function with $Y^a$ as its argument.
First, note that there is a vector associated to $\lambda_a$:
\[
\lambda^a:=\Omega^{ab}\lambda_b
\]
so we can write
\be
F_\lambda(Y)=\Omega_{ab}\lambda^aY^b=\Omega(\lambda,Y)
\ee
If we are now given another label $\nu$, such that $G_\nu(Y)=\nu_aY^a$,
we can compute the Poisson Bracket
\be
\{F_\lambda,G_\nu\}=\Omega^{ab}\nabla_aF_\lambda(Y)\nabla_bG_\nu(Y)=
\Omega^{ab}\lambda_a\nu_b
\ee
Since the two-form is non-degenerate we can re-write it as
$\{F_\lambda,G_\nu\}=\Omega_{ab}\lambda^a\nu^b$. Thus,
\be
\{\Omega(\lambda,Y),\Omega(\nu,Y)\}=\Omega(\lambda,\nu)
\ee
As we shall see in Sec.~\ref{C.2} we can also make such a selection
of linear functions for the Maxwell field.

The quantum representation is the ordinary Schr\"odinger representation
where the Hilbert space is $\H=L^2(\R^3,\d^3 x)$ and the operators
are represented:
\be
(\hat{1}\cdot\Psi)(q)=\Psi(q)\qquad(\hat{q^i}\cdot\Psi)(q)=q^i\Psi(q)
\qquad(\hat{p}_i\cdot\Psi)(q)=-i\hbar\frac{\partial}{\partial q^i}\Psi(q)
\ee
Thus, we recover the conventional quantum theory.

\section{Classical description for the Maxwell Field}
\label{C.1}

In the classical phase space description of the Maxwell field there are
two equivalent but complementary viewpoints, namely the {\it covariant}
and the {\it canonical} formalisms. In what follows we shall develop
both approaches and show their equivalence.

\subsection{Covariant Phase Space}
\label{C.1.1}

In this part we shall introduce and employ  the {\it covariant
phase space formulation} \cite{covariant}.
Since in our opinion this formalism is not
widely known, we shall outline the main steps using the Maxwell
field as an example.
The starting point for the construction of the covariant phase space
is the identification of the symplectic vector space $\Gamma$, the
phase space of the problem, starting from 
solutions to the equations of motion. Let us start by writing down
the action for the free Maxwell theory:
\ba
S_{\rm M}&:=&-{\frac{1}{4}}\int_M F^{ab}F_{ab}\;\sqrt{|g|}\,\d^4\! x\, ,
\nonumber\\
&=&- {\frac{1}{2}}\int_M F^{ab}\nabla_{[a}A_{b]}\;\sqrt{|g|}\,
\d^4\! x\, .
\label{action}
\ea
where $F_{ab}:= 2\nabla_{[a}A_{b]}$. 
The variation of the action is given by,
\begin{equation}
\delta S_{\rm M}=\int_M(\nabla_a F^{ab})\delta A_a\;\sqrt{|g|}\,\d^4\! x-
\int_{\partial M}F^{ab}\delta A_b\;\d\Sigma_a\, .\label{variation}
\end{equation}
The volume term tells us that the action is extremized when
$\nabla_aF^{ab}=0$. Since we are assuming that there exists
a one-form $A_a$ such that its exterior derivative is the Maxwell field
$F_{ab}:=2\nabla_{[a}A_{b]}$, the equation $\nabla_{[a}F_{bc]}=0$
is automatically satisfied (the Bianchi identity).
 Therefore we have the full set of Maxwell
equations. The second term in Eq (\ref{variation}), the boundary term,
is often referred to as the {\it symplectic current}. It can be
interpreted as a 1-form on the space $\bar{\Gamma}$ of solutions to
the equations of motion (it is analog to the symplectic potential $\omega$
introduced in Sec.~\ref{sec0}).
 It is acting on the vector $\delta A_a$ and
producing a number. We can take now another `variation' of this
term in order to get the conserved (pre)-symplectic structure
$\w(\cdot,\cdot)$,
\begin{equation}
\w(\delta A, \widetilde{\delta A}):= \int_{\Sigma}(\delta F^{ab}
\widetilde{\delta A_b}-\widetilde{\delta F}^{ab}\delta A_b)
\; \,\d\Sigma_a\, ,
\end{equation}
where $\Sigma$ is any Cauchy surface in the space-time $M$\footnote{
A Cauchy surface is a space-like surface $\Sigma$ whose domain of dependence
in the entire space-time $M$.}.
We have
not been very precise about functional analytic issues. We are just 
requiring falloff conditions (on any $\Sigma$) such that the symplectic
form at spatial $\infty$ vanishes. If, in particular, we
restrict ourselves to solutions of the Maxwell equations that induce
data of {\it compact support}
 on any Cauchy surface, that conditions will
be satisfied\footnote{A function of compact support is
a function that vanishes outside a compact region of $\Sigma$.}. 
This bilinear mapping defined by $\w$ is, however, degenerate. There are
tangent vectors $X_\a$ such that $\w(X,Y)=0, \;\forall \;
Y\in T\bar{\Gamma}$\footnote{We denote by $X_\a$ the infinite dimensional
tangent vector (with abstract index $\a$) defined by $X_a(x)$.}.
These are the degenerate directions of $\w$. The fact that the two-form
$\w$ is degenerate on $\bar{\Gamma}$ is an indication that there is some
{\it gauge freedom} in the system. Let us now try to identify what the
degenerate directions of $\w$ are.
Since we are restricting ourselves to the space $\bar{\Gamma}$,
the tangent vectors satisfy the {\it linearized} equation of motion,
that in this case coincide with the Maxwell equations.
Consider vectors of the type $X_a=\nabla_a \Lambda$ for some function
$\Lambda$. Then, using the fact that it satisfies $\nabla_{[a}X_{b]}=0$
we have,
\ba
\w(X,\delta A)& = &\int_\Sigma-\delta F^{ab}\nabla_b\Lambda\; 
\,\d\Sigma_a\nl
&=& \int_\Sigma\Lambda \nabla_b(\delta F^{ab})\; 
\,\d\Sigma_a=0\, .\nonumber
\ea
We can conclude that the degenerate directions of $\w$ are of the form
$\nabla_a\Lambda$. This is the manifestation, in the covariant
phase space approach, of the ``gauge freedom'' present in
electro-magnetism.
 In order to get a
true symplectic space, we should take the quotient of $\bar{\G}$
by the degenerate directions of $\w$  to get $\Gamma$, the (reduced)
phase space of the theory. Note that $\G$ can be equivalently
parameterized by the equivalence class of gauge potentials $[A_a]$,
where $A\sim\bar{A}$ iff $A_a=\bar{A}_a +\nabla_a\Lambda$, or alternatively,
by the gauge fields $F_{ab}$, satisfying Maxwell equations.

We can now write the (weakly non-degenerate) symplectic form on $\G$:
\begin{equation}
\w(F,\widetilde{F})=\int_\Sigma(F^{ab}\widetilde{A}_b-
\widetilde{F}^{ab}A_b)\;
 \,\d\Sigma_a\label{w(f,f)}\, .
\end{equation}
Note that it is well defined on $\G$ since it does not depend on the
representative of the equivalence class $[A]$. Note that in writing
(\ref{w(f,f)}) we have used the fact that $\G$ is a linear space
and therefore we can identify points in $\G$ with tangent vectors.

The next step is to construct observables of the theory, namely,
real valued functions on $\G$. A natural strategy is to use
the symplectic form in order to construct such functions. Let $h_a$ be
a ``test 1-form''. The observable ${\cal O}[h]:\G\rightarrow \R$, labeled
by $h$, is defined in complete analogy with Sec.~\ref{sec0} by the
expression,
\begin{equation}
({\cal O}[h])(F):=\w(F,T)=\int_\Sigma (F^{ab}h_b-T^{ab}A_b)\; 
\,\d\Sigma_a\, ,\label{observ}
\end{equation}
where $T_{ab}:=2\nabla_{[a}h_{b]}$. We need it to be a well defined
function on $\G$, so ${\cal O}[h]$ should be invariant under gauge
transformations $A_a\rightarrow A_a+\nabla_a\Lambda$. Thus, we have
to require that
\begin{equation}
\int_\Sigma T^{ab}\nabla_b\Lambda\; \,\d\Sigma_a=0\, ,
\end{equation}
which implies $\nabla_aT^{ab}=0$.  Therefore,
 an element $h_a$ of $\G$ defines by
itself  a linear observable, since $h_a$ and $h_a+\nabla_a\Lambda$ define the
same function. In the quantum theory, to each
of this observables there will correspond a quantum operator,
making the correspondence between solutions to Maxwell equations
and  quantum operators precise. 

Let us re-write the symplectic form (\ref{w(f,f)}) in terms of the
familiar electric and magnetic fields. Recall that given a local
observer with four velocity $t^a$ ($t_at^a=-1$), then the {\it
electric field} with respect to this observer is given by
$E_a:=t^bF_{ba}$. It is naturally defined as a 1-form. Since we 
have a metric we can `raise' the index and define the corresponding
vector field. We can also define the {\it dual tensor} of the 
field $F_{ab}$ by: ${}^*\!F^{ab}:={\T\frac{1}{2}}\epsilon^{abcd}F_{cd}$,
where $\epsilon^{abcd}$ is the canonical volume form defined
by the metric
$g_{ab}$ with all its indices raised with the metric. The magnetic
field is defined by $B_a:=t^b {}^*\!F_{ba}$. In the integrand of the
symplectic form, one is contracting the tensor $F^{ab}$ with the
unit normal $n_a$ to the surface $\Sigma$ (that is the meaning of
$\d\Sigma_a:=\epsilon_{abcd}\,\d\Sigma^{bcd}$), so we get naturally the
 electric field $E^a$ with respect to $\Sigma$. 
We can now express (\ref{w(f,f)}) as follows,
\begin{equation}
\w (F,\tilde{F});=\int_\Sigma(E^a\tilde{A}_a-\tilde{E}^aA_a)\;\sqrt{h}
\,\d^3\! x\, .
\end{equation}
This expression can be rewritten in terms of objects defined purely
on the hyper-surface $\Sigma$. We can write,
\ba
F^{ab}A_b\,\d\Sigma_a &=&
\frac{1}{2}\,\epsilon^{abcd}{}^*\!F_{cd}\,A_b\,\d\Sigma_a\, ,\nl
&=&\frac{1}{2}\,{}^*\!F_{cd}\,A_b\,\epsilon^{abcd}\epsilon_{afgh}\,\d
\Sigma^{fgh}\, ,\nl
&=&-\frac{1}{2}\,{}^*\!F_{cd}\,A_b\,\d\Sigma^{cdb}\, .\nonumber
\ea
Therefore, one can take the 3-form ${}^*\!F\wedge A$ and
integrate it on $\Sigma$,
\begin{equation}
\w (F,\tilde{F})=-\frac{1}{2}\int_\Sigma ({}^*\!F_{[ab}\tilde{A}_{c]}-
{}^*\!\tilde{F}_{[ab}{A}_{c]})\;\d\Sigma^{abc}.
\end{equation}
Note that the pullback to $\Sigma$ of the dual tensor ${}^*\!F_{ab}$
is, in a 3-dimensional sense, the electric field two-form:
$E_{ab}:= {}^*\!F_{ab}$. This is naturally dual to a vector density
of weight one $\tilde{E}^c:=\tieta^{cab}E_{ab}$, which is, as we
shall later see, the
electric field arising from the canonical approach. 
Here, $\tieta^{abc}$ is the naturally defined completely anti-symmetric
Levi-civita density of weight one on $\Sigma$.

Finally, one can ask what the Poisson Bracket of the observables 
defined by (\ref{observ}) is. Given $h_a$ and ${h}^\prime_a$ in $\G$ the
Poisson bracket of the observables they define is given by,
\begin{equation}
\{{\cal O}[h],{\cal O}[{h}^\prime]\}:=\w(T,{T}^\prime)=
\int_\Sigma(T^{ab}{h}^\prime_b-
{T}^{\prime ab}h_b)\; \,\d\Sigma_a\, .
\end{equation}

We have seen that starting from the action, there is a naturally defined
symplectic structure $\w$ on $\G$. We constructed the lineal observables
${\cal O}[h]$, the generators of the algebra $\S$ and computed the
Poisson bracket amongst them.
We shall now go to the canonical approach.

\subsection{Canonical Phase Space}
\label{C.1.2}

In this part we shall present the canonical phase space description
of the Maxwell Field, which is normally known as the `Dirac Analysis'
\cite{dirac}.
However, our presentation will be `covariant' in the sense that
our analysis is coordinate free; that is, we
do not assume any coordinate system on $M$. 
The action (\ref{action}) can be written in a $3+1$ fashion. First we
write the expression for the action as follows,
\begin{equation}
S = -{\frac{1}{4}}\int_M g^{ab}g^{cd}F_{ac}F_{bd}\;\sqrt{|g|}\,
\d^4\! x\label{action2}
\end{equation}
Next, we decompose the space-time metric as follows:
$g^{ab}=h^{ab}-n^an^b$. Here $h^{ab}$ is the (inverse of) the induced metric
on the Cauchy hyper-surface $\Sigma$ and $n^a$ the unit normal to
$\Sigma$. We also introduce an everywhere time-like vector field $t^a$
and a `time' function $t$ such that the hyper-surfaces $t=$constant
are diffeomorphic to $\Sigma$ and such that $t^a\nabla_at=1$. We can
write $t^a=Nn^a+N^a$. The volume element is given by $\sqrt{|g|}=N\,\sqrt{h}$.
Using this identities in Eq.(\ref{action2}) we get,
\ba
S&=& -{\frac{1}{4}}\int_I\d t\int_\Sigma N\sqrt{h}\,\Big\{ h^{ac}h^{bd}
F_{ab}F_{cd}-\nl
&{}&{\frac{2}{N^2}}h^{ac}\left[(\Lt A_a-\nabla_a(t\cdot A)+N^bF_{ab})
(\Lt A_c-\nabla_c(t\cdot A)+N^dF_{cd})\right]\Big\}\, ,\label{ac3+1}
\ea
where $(t\cdot A):=t^bA_b$, and $I=[t_0,t_1]$ is an interval in the
real line.
 Note that since for all the terms
in the previous equation, both
the one-form $A_a$ and the field strength $F_{ab}$ are contracted
with purely ``spatial'' objects ($n^aN_a=n^ah_{ab}=0$),
then both $A_a$ and $F_{ab}$ in
(\ref{ac3+1}) are the pull-backs to $\Sigma$ of the space-time objects.
For simplicity, we shall continue to write $A_a$ for the 3-dimensional
potential.

From the $3+1$ form of the action (\ref{ac3+1}) we can find the
momenta canonically conjugated to $A_a$:
\begin{equation}
\tilde{\Pi}^a:=\frac{\delta S}{\delta (\Lt A_a)}={\T\frac{\sqrt{h}}{N}}
h^{ac}(\Lt A_c-\nabla_c(t\cdot A)+N^dF_{cd})\label{pi=some}\, .
\end{equation}
It can be rewritten as,
\begin{equation}
\tilde{\Pi}^a={\T\frac{\sqrt{h}}{N}}h^{ac}(t^b-N^b)F_{bc}=
{\T\frac{\sqrt{h}}{N}}
h^{ac}Nn^bF_{bc}=\sqrt{h}\,E^a\, ,
\end{equation}
thus, the canonically conjugated momenta is just the {\it densitized}
electric field (w.r.t. $\Sigma$). In this subsection, a `tilde' over
a tensor means that it is a density of weight one.

The Eq.(\ref{pi=some}) can be solved for the `velocity', $\Lt A_a$,
\begin{equation}
\Lt A_a={\T\frac{N}{\sqrt{h}}}\,h_{ac}\tP^a+\nabla_c(t\cdot A)-N^dF_{cd}
\end{equation}
We can perform a Legendre transform of the Lagrangian density in order to
find the Hamiltonian:
\ba
H &:=& \int_\Sigma\d^3\! x\,\left(\tP^a\Lt A_a-\tilde{{\cal L}}\right)\nl
&=& \int_\Sigma \d^3\! x\,\Big(-(t\cdot A)\nabla_a\tP^a-
N^dB_{ad}\tP^a+{\T\frac{N}{2\sqrt{h}}}h_{ac}\tP^a\tP^c+
{\T\frac{N\sqrt{h}}{4}}h^{ac}h^{bd}B_{ab}B_{cd}\Big)\, .\label{hamil}
\ea
We have denoted by $B_{ab}=\underline{F_{ab}}$ the field strength of
the 3-dimensional potential $A_a$.
 It is related to the magnetic field in the following
way: $B^a:=\frac{1}{\sqrt{h}}\tilde{\eta}^{abc}B_{bc}$. The last term
in  (\ref{hamil}) can be rewritten: $h^{ac}h^{bd}B_{ab}B_{cd}=B^eB^f
{\epsilon^{cd}}_e\epsilon_{cdf}=2h_{ab}B^aB^b$.
In the `Dirac analysis' of the action (\ref{action2}) the first step
is to identify the {\it configuration variables}. In this case, these
are pairs $(\phi:=(t\cdot A), A_a)$, that is, we have four configuration
degrees of freedom per point.
 In the action there is no
term corresponding to time derivative of $\phi$ so we have a primary
constraint $\chi_1=\tilde{\Pi}_\phi\approx 0$.
The basic Poisson brackets are,
\begin{equation}
\{A_a(x),\tP^b(y)\}=\delta^b_a\delta^3(x,y)\quad;\quad
\{\phi(x),\tilde{\Pi}_\phi(y)\}=\delta^3(x,y)\, .\label{mxwpb}
\end{equation}
Asking that the constraint be preserved in time with respect to the
Hamiltonian (\ref{hamil}) leads to the secondary constraint
$\chi_2:=\nabla_a\tP^a\approx 0$. There are no extra constraints. They form
a {\it First Class} system\footnote{A first class system has the property
that the Hamiltonian vector fields $X^\alpha_{\chi_1}$ and
$X^\alpha_{\chi_2}$ are tangent to the $\chi_1=\chi_2=0$ surface.}.
One can eliminate the first one by
giving the gauge condition $\chi_3:=\phi-\lambda(\x)\approx 0$,
with $\lambda$
an arbitrary function on $\Sigma$. We can reduce the constraints
$(\chi_1,\chi_3)$ since they form a second class pair. We are then left
with the Gauss constraint $\chi_2=\nabla_a\tP^a\approx 0$. Now, $\phi$ has
the role of a {\it Lagrange multiplier}. Therefore, the phase space
$\Gamma^\prime$ is coordinatized by the pairs $(A_a,\tP^b)$, having
three degrees of freedom per point. The
constraint surface $\hat{\G}$ are the point in $\G^\prime$ where the
Gauss constraint is satisfied. In the canonical picture, gauge
transformations are those canonical transformations generated by
the (first class) constraints.
The reduced phase space $\G_{\rm c}$ is then the
space of orbits generated by the gauss constraint in $\hat{\G}$.
The canonical transformation generated by the (smeared)
Gauss constraint, $G[\lambda]=\int_\Sigma\lambda\nabla_b\tP^b\d^3\! x$,
is given by,
\begin{equation}
A_a\longrightarrow A_a-\nabla_a\lambda\, .
\end{equation}
Therefore, the (reduced) phase space is given by pairs $([A], \tP)$ of
gauge equivalence class of connections and vector densities satisfying
Gauss' law. Thus, we recover the two {\it true degrees of freedom}  the
the Maxwell field has (corresponding to the two types of polarization).
One alternative to the reduced phase space description is to impose
a {\it gauge condition}
in order to select one particular representative from the equivalence
class. A convenient gauge choice in this case is to ask that
$\chi_4:=\nabla^aA_a=0$.
This is a good gauge condition since the pair $(\chi_2,\chi_4)$ forms
a second class pair\footnote{A second class pair of constraints is such that
the symplectic structure restricted to the surface they define is
non-degenerate.}.
Thus,
we can coordinatize $\G_{\rm c}$ by
$(A_a,E^a)$, a pair of divergence-less (transverse) vector fields on
$\Sigma$. We have used the fact that we have a metric on $\Sigma$
to de-densitize the momenta $\tP$. 
 
The Poisson brackets (\ref{mxwpb}) induce a (weakly) non-degenerate
symplectic form $\w$ on pairs of tangent vectors $(\delta A,\delta E)$ on
$T^*\G^\prime$:
\begin{equation}
\w\left( (\delta A,\delta E);({\delta A}^\prime,{\delta E}^\prime)
\right) =\int_\Sigma \sqrt{h}\,\d^3\! x 
\left({\delta A}^\prime_a
\delta E^a- \delta{A}_a{\delta E}^{\prime a} \right)\, .
\end{equation}

The Poisson Brackets on transverse traceless quantities (The Dirac
bracket in the standard terminology) are given by,
\be
\{A_a^{\rm T}(x),E_{\rm T}^b(y)\}=\delta^a_b\delta^3(x,y)-\Delta^{-1}
D^bD_a\delta^3(x,y),
\ee
where $\Delta$ is the Laplacian operator compatible with the metric
$h_{ab}$.

We can now relate the two approaches and see that the phase space
$\G$ from last section is precisely the space $\G_{\rm c}$
constructed via the canonical approach. The key observation is that
there is a one to one correspondence between a pair of initial data of 
compact support on $\Sigma$, satisfying the transverse condition,
and solutions to the Maxwell equations on $M$, modulo gauge
transformations (an element of $\Gamma$)  \cite{wald1}.
Therefore, to each element $F_{ab}$ in $\G$ there is a pair
$(A_a, E^a)$ on $\G_{\rm c}$ ($2\nabla_{[a}A_{b]}=\underline{F_{ab}}$
and $E^a=h^{ab}n^cF_{cb}$ and more importantly, for each pair, there
is a solution to Maxwell's equations that induces the given initial
data on $\Sigma$. Here, `underline' denotes restriction to $\Sigma$.
From now on, we shall refer to elements of the vector space $\G$
in-distinctively  either as $F_{ab}$ or as $(A_a,E^b)$.

Observables for the space $\G$ can be constructed directly by giving
smearing functions on $\Sigma$ (compare to the discussion of the previous
section in which the observables were constructed from {\it space-time}
smearing objects). Given a 1-form $g_a$ on $\Sigma$ we can define,
\begin{equation}
E[g]:=\int_\Sigma\sqrt{h}\;\d^3\! x\,E^ag_a\, .\label{observ2}
\end{equation}
Similarly, given a vector field $f^a$ we can construct, 
\begin{equation}
A[f]:=\int_\Sigma\sqrt{h}\;\d^3\! x\,A_af^a\, ,\label{observ3}
\end{equation}
Asking that $E[g]$ be gauge invariant does not impose any condition on
$g_a$, since  Gauss' law does not `move' the electric field. Note
however that $E[g]$ takes the same value for $g_a$ and $g_a+\nabla_a\lambda$.
It is convenient to restrict ourselves to $g_a$ satisfying $\nabla^ag_a=0$.
 The requirement that $A[f]$ be gauge invariant tells us 
that $\nabla_af^a=0$. Therefore, in order to get well defined operators,
we need the pairs $(g_a,f^b)$ to belong to the phase space $\G$. These
are the precise images  of the observables (\ref{observ}) given by
the identification of phase spaces. The  relation is given by
$g_a=\underline{h_a}$ and $f^a=2\nabla^{[a}h^{b]}n_b$.

Note that any pair of test fields $(g_a,f^a)\in \G$ 
defines a linear observable,
but they are `mixed'. More precisely, a connection $g_a$ in $\Sigma$, that
is, a pair $(g_a,0)\in \G$
gives rise to an {\it electric field} observable $E[g]$ and,
conversely, a vector field $(0,f^a) \in \G$ defines a {\it connection}
observable $A[f]$.

As we have seen, the phase space $\G$ can be alternatively described
by equivalence classes of solutions to the Maxwell Equations in the
covariant formalism or by pairs of transverse vector fields on a
Cauchy surface $\Sigma$ in the canonical approach. In both cases, the
elements of the algebra $\S$ to be quantized are linear functionals of
the basic fields. In the covariant case
they are constructed out of space-time smearing fields and in the
canonical language out of a pair of space smearing fields. 
In the next section we consider the construction of the quantum theory.

\section{Quantization}
\label{C.2}

In this section we shall construct the quantum theory.
This section is divided into four parts. In the first one we construct
the one-particle Hilbert space $\H$ from the phase space $\G$
of the classical theory.
In the second part, we introduce the symmetric Fock space $\F$
associated with the one-particle Hilbert Space $\H$. In the third
part we find representations of the CCR an the given Fock space. Finally,
in the last part we give some examples.   

\subsection{One-particle Hilbert Space}
\label{C.2.0}

The first step in the quantization program is to identify the 1-particle
Hilbert space $\H$.
 The strategy
is the following: start with $(\G,\w)$ a symplectic vector space
and define $J:\G\rightarrow\G$, a linear operator such that $J^2=-1$. The
{\it complex structure} $J$ has to be compatible with the symplectic
structure. This means that the bilinear mapping defined by
$\mu(\cdot,\cdot):=\w(\cdot,J\cdot)$ is
a positive definite metric on $\G$.
The Hermitian (complex) inner product is then given by,
\begin{equation}
\langle\cdot,\cdot\rangle=\T\frac{1}{2\hbar}\mu(\cdot,\cdot)+
i\T\frac{1}{2\hbar}\w(\cdot,\cdot)\label{<.,.>}\, .
\end{equation}
 The complex structure $J$ defines a a natural splitting of $\G_\C$,
the complexification of $\G$, in the following way: Define the
`positive frequency' part to consist of vectors of the form
$\Phi^+:=\frac{1}{2}(\Phi-iJ\Phi)$ and the `negative frequency' part
as $\Phi^-:=\frac{1}{2}(\Phi+iJ\Phi)$. Note that $\Phi^-=\overline{\Phi}^+$
and $\Phi=\Phi^++\Phi^-$. Since $J^2=-1$, the eigenvalues of $J$ are
$\pm i$, so one is decomposing the vector space $\G$ in eigenspaces of
$J$: $J(\Phi^{\pm})=\pm i\Phi^{\pm}$. We have used the term `positive
frequency' since in the case of $M$ Minkowski space-time  that is the
standard decomposition. The Hilbert space $\H$ is the completion of
$\G$ with respect to the inner product (\ref{<.,.>}).

There are two alternative but completely equivalent description of the
1-particle Hilbert space $\H$:
\begin{enumerate}

\item $\H$ consists of {\it real} valued functions (solution to the
Maxwell equation for instance), equipped with the complex structure $J$.
The inner product is given by (\ref{<.,.>}). 

\item $\H$ is constructed by complexifying the vector space $\G$ (tensoring
with the complex numbers) and then decomposing it using $J$ as described
above. In this construction, the inner product is given by,
\begin{equation}
\langle\Phi,\tilde{\Phi}\rangle=\T\frac{i}{\hbar}\w(\Phi^-,\tilde{\Phi}^+)
\end{equation}
Note that in this case, the 1-particle Hilbert space consists of
`positive frequency' solutions.

\end{enumerate}

It is important to note that the only input we needed in order to
construct $\H$ was the complex structure $J$. For a general space-time
there is no preferred one. This in turn leads to the infinite ambiguity
in the representation of the CCR. In the case of stationary space-times
there is a preferred, canonical, complex structure given by the 
Killing field. 
This construction for the case of the Klein Gordon field is
described in \cite{6}. For Minkowski space-time there are several
ways of characterizing the usual quantization. The standard textbook
treatment uses a (globally inertial) time coordinate $t$ to perform the
positive-frequency decomposition. Another way of selecting this
decomposition is to ask that the vacuum on the resulting theory be
Poincar\'e invariant. A third way is to ask that the coherent states in
the quantum theory have the same energy as the classical solution on
which they are peaked \cite{aa:am80}.

\subsection{Fock Space}
\label{C.2.1}

Given a Hilbert space $\H$ there is a natural way of constructing
its associated Fock Space. In this part
we shall describe this universal construction of the Fock space
associated to the Hilbert space $\H$ and then give in detail the
representation for the Maxwell field in Minkowski space-time.

The {\it symmetric Fock space} associated to $\H$ is defined
to be the Hilbert space
\begin{equation}
\F_{\rm s}(\H):=\bigoplus^{\infty}_{n=0}
\left(\bigotimes^n{}_{\rm s}\H\right)\, ,
\end{equation}
where we define the {\it symmetrized tensor product} of $\H$, denoted
by $\bigotimes^n{}_{\rm s}\H$, to be the subspace of the n-fold tensor
product ($\bigotimes^n\H$), consisting of totally symmetric maps
$\alpha:\overline{\H}_1\times\cdots\times\overline{\H}_n\rightarrow \C$
satisfying
\begin{equation}
\sum \left|\alpha(\bar{e}_{i_1},\ldots,\bar{e}_{i_n})\right|^2<\infty\, .
\end{equation}
The Hilbert space $\overline{\H}$ is the {\it complex conjugate} of
$\H$ with $\{\bar{e}_{1},\cdots,\bar{e}_{j},\cdots\}$ an orthonormal
basis. We are also defining $\bigotimes^0\H=\C$.

We shall introduce the abstract index notation for the Hilbert spaces
since it is most convenient way of describing the Fock space. Given
a space $\H$, we can construct the spaces $\overline{\H}$, the
complex conjugate space; $\H^*$, the {\it dual space}; and
$\overline{\H}^*$ the dual to the complex conjugate. In analogy with
the notation used in spinors, let us denote elements of $\H$ by
$\phi^A$, elements of $\overline{\H}$ by $\phi^{A^\prime}$. Similarly,
elements of ${\H}^*$ are denoted by $\phi_A$ and elements of
$\overline{\H}^*$ by $\phi_{A^\prime}$. However, by using Riesz lemma,
we may identify $\overline{\H}$ with $\H^*$ and $\H$ with
$\overline{\H}^*$. Therefore we can eliminate the use of primed indices,
so $\overline{\phi}_{A}$ will be used for an element in
$\overline{\H}^*$ corresponding to the element $\phi^A\in \H$.
An element $\phi\in\bigotimes^n{}_{\rm s}\H$ then
consists of elements satisfying
\begin{equation}
\phi^{A_1\cdots A_n}=\phi^{(A_1\cdots A_n)}
\end{equation}
An element $\psi\in \bigotimes^n\overline{\H}$ will be denoted as
$\psi_{A_1\cdots A_n}$. In particular, the inner product of vectors
$\psi,\phi\in \H$ is denoted by
\begin{equation}
\langle\psi,\phi\rangle=:\overline{\psi}_A\phi^A
\end{equation}

A vector $\Psi\in\F_{\rm s}(\H)$ can be represented, in the abstract index
notation as
\begin{equation}
\Psi=(\psi,\psi^{A_1},\psi^{A_1A_2},\ldots ,\psi^{A_1\ldots A_n}, \ldots )\, ,
\end{equation}
where, for all $n$, we have $\psi^{A_1\ldots A_n}=\psi^{(A_1\ldots A_n)}$.
The norm is given by
\begin{equation}
|\Psi|^2:=\overline{\psi}\psi+\overline{\psi}_A\psi^A+\overline{\psi}_{A_1A_2}
\psi^{A_1A_2}+\cdots < \infty\, .
\end{equation}
Now, let $\xi^A\in\H$ and let $\overline{\xi}_A$
denote the corresponding element
in $\overline{\H}$. The {\it annihilation operator} ${\cal A}(\bar{\xi}):
\F_{\rm s}(\H)\rightarrow \F_{\rm s}(\H)$ associated to
$\overline{\xi}_A$ is denoted by
\begin{equation}
{\cal A}(\bar{\xi})\cdot\Psi:=(\overline{\xi}_A\psi^A,\sqrt{2}\,
\overline{\xi}_A
\psi^{AA_1},\sqrt{3}\,\overline{\xi}_A\psi^{AA_1A_2},\ldots )\, .\label{crea}
\end{equation}
Similarly, the {\it creation operator} ${\cal C}({\xi}):
\F_{\rm s}(\H)\rightarrow \F_{\rm s}(\H)$ associated with
$\xi^A$ is defined by
\begin{equation}
{\cal C}(\xi)\cdot\Psi:=(0,\psi\xi^{A_1},\sqrt{2}\,\xi^{(A_1}\psi^{A_2)},
\sqrt{3}\,\xi^{(A_1}\psi^{A_2A_3)},\ldots )\, .\label{aniqui}
\end{equation}
If the domains of the operators are defined to be the subspaces of
$\F_{\rm s}(\H)$ such that the norms of the right sides of eqs. (\ref{crea})
and (\ref{aniqui}) are finite then it can be proven that
${\cal C}(\xi)=({\cal A}(\bar{\xi}))^\dagger$. It may also be verified that
they satisfy the commutation relations,
\begin{equation}
\left[{\cal A}(\bar{\xi}),{\cal C}(\eta)\right]=\bar{\xi}_A\eta^A\,{\rm I}\, .
\label{CCR2}
\end{equation}
A more detailed treatment of Fock spaces can be found in 
\cite{geroch,wald2,reeds}.

\subsection{Representation of the CCR}
\label{C.2.2}

In the previous section we saw that we could construct linear observables
in $(\G,\w)$, in either of the classical constructions.
 For the covariant picture
the observables are given by (\ref{observ}) and in the canonical by
(\ref{observ2}) and (\ref{observ3}). This is the set ${\cal S}$ of
observables for which there will correspond a quantum operator. Thus,
for ${\cal O}[h]\in {\cal S}$ there is an operator $\hat{\cal O}[h]$.
We want the {\it Canonical Commutation Relations} to hold,
\begin{equation}
\left[\hat{\cal O}[h],\hat{\cal O}[\tilde{h}]\right]=i\hbar
\{{\cal O}[h],{\cal O}[\tilde{h}]\}=i\hbar\,\w(h,\tilde{h})\, .
\end{equation}
Then we should find a Hilbert space and a representation thereon of
our basic operators satisfying the above conditions. We have all the
structure needed at our disposal. Let us take as the Hilbert space
the symmetric Fock space $\F_{\rm s}(\H)$ and let the operators be
represented as
\begin{equation}
\hat{\cal O}[h]\cdot\Psi:=\hbar\left( {\cal C}(h) + 
{\cal A}(\overline{h})\right)
\cdot \Psi\, .
\end{equation}
Let us denote by $h^A$ the abstract index representation
corresponding to $h_a$ in $\H$.
First, note that by construction the operator is self-adjoint.
It is straightforward to check that the commutation relations are
satisfied,
\ba
\left[\hat{\cal O}[h],\hat{\cal O}[{h}^\prime]\right]&=&
\hbar^2\big[{\cal C}[h],{\cal A}[\overline{h}^\prime]\big]+
\hbar^2\big[{\cal A}[\overline{h}],
{\cal C}[{h}^\prime]\big]\nl
&=&\hbar^2\,( \overline{h}_Ah^{\prime A}-\overline{h}^\prime_Ah^A)\nl
&=&\hbar^2\,(\langle h, h^\prime\rangle-\langle h^\prime,h\rangle)\nl
&=& 2i\hbar^2\,{\rm Im}(\langle h,h^\prime\rangle)=i\hbar\,\w(h,h^\prime)\, ,
\ea
where we have used (\ref{CCR2}) in the second line
 and (\ref{<.,.>}) in the last line. Note that in this last calculation
we only used general properties of the Hermitian inner product and
therefore we would get a representation of the CCR for {\it any} 
inner product $\langle\cdot,\cdot\rangle$. Since the inner product is
given in turn by a complex structure $J$, we see that there is a one to one
correspondence between them.

\subsection{Examples}
\label{C.2.3}

As mentioned at the end of Sec.~\ref{C.2.0}, the choice of a complex
structure $J$ is far from being a straightforward process. For a general
space-time, there is no a-priori criteria to select one. Furthermore,
there are an infinite number of choices that give {\it inequivalent} quantum
theories \cite{wald2}. In the special case that there exists a time-like
Killing vector field $t^a$ on the spacetime $(M,g)$; that is,
for a stationary space-time,
there exists a canonical choice of complex structure given by the
killing field. From the physical viewpoint, this choice is motivated
because it gives to coherent states peaked at a particular solution
an energy equal to the classical energy associated to that solution
\cite{aa:am80}. The complex structure is given by,
\be
J:=-(-\Lt\cdot \Lt)^{-1/2}\,\Lt
\ee

A particular important example of a space-time with a globally defined
Killing field is Minkowski space-time
(in fact it has an infinite number of such
vector fields, one for each inertial reference frame). From now on,
let us restrict our attention to Minkowski space-time and inertial
hyper-surfaces $\Sigma$. Therefore, the induced metric $h_{ab}$ is the
Euclidean flat metric. We will perform two different decompositions
of $\G$, for two different complex structures. First, we shall consider
the ordinary `positive frequency' decomposition. This leads to the
standard quantum theory of the free Maxwell field found in textbooks.
Next, we decompose $\G$ in self-dual and anti-self-dual fields.


\subsubsection{Positive Frequency Decomposition}

Since it is completely equivalent to use the covariant or canonical
notation, we shall denote elements of $\G$ as
pairs $(A^{\rm T}_a,E_{\rm T}^a)$, of transverse (i.e. divergence-free)
vector fields.  
The first step in the quantization is the introduction
of the complex structure $J:\G\rightarrow \G$.   
It is given by,
\begin{equation}
J\cdot\left(A_a\atop E_a\right):=\left(-\Delta^{ 1/2}E_a\atop
\Delta^{ -1/2}A_a\right)\, .\label{compst}
\end{equation}
Next, we can construct the {\it projector operator} $K^+:\G\rightarrow\G_\C$,
such that $F^+_{ab}=K^+(F_{ab})$ is the {\it positive frequency} part
of $F_{ab}\in \G$. The projector is given by the following action in terms
of the pairs of initial data,
\begin{equation}
K^+\cdot\left(A_a\atop E_a\right):=\frac{1}{2}\left(A_a-i\Delta^{ -1/2}
E_a \atop E_a+i\Delta^{ 1/2}A_a\right)\, .
\end{equation}
With this definitions, we can construct the inner product in $\H$. For
$F,\tilde{F}$ in $\H$ we have,
\ba
\langle F,\tilde{F}\rangle &=& \frac{i}{\hbar}\w(\overline{F}^+,\tilde{F}^+)\nl
&=& \frac{i}{\hbar}\int_{\Sigma}\d^3\! x
(\overline{E}^{+a}\tilde{A}^+_a-\tilde{E}^{+a}\overline{A^+}_a)\nl
&=&\frac{i}{4\hbar}\int_{\Sigma}\d^3\! x\,\big[(E^a\tilde{A}_a-\Delta^{1/2}
A^a\Delta^{-1/2}\tilde{E}_a-\tilde{E}^aA_a+\Delta^{1/2}\tilde{A}^a
\Delta^{-1/2}E_a)\nl
&{}& -i(\tilde{A}_a\Delta^{1/2}A^a+E^a\Delta^{1/2}\tilde{E}_a
+A_a\Delta^{1/2}\tilde{A}^a+\tilde{E}^a\Delta^{-1/2}E_a)\big]\, .
\ea
The norm  of $(g_a,f^a)\in \H$ is given by,
\begin{equation}
\langle(g,f),(g,f)\rangle=\frac{1}{2\hbar}\int_{\Sigma}\d^3\! x\,
\big(g_a\Delta^{1/2}g^a+f^a\Delta^{-1/2}f_a)\big)\, .\label{norm3}
\end{equation}
One should keep in mind that all the objects $(g_a,f^a)$ are transverse.
The reason for this requirement is that the complex structure takes a
very simple form (\ref{compst}) in terms of transverse vector fields,
making also the expression for the norm look simple (\ref{norm3}).

We are now in position of asking whether an observable generated by
the pair $(g_a,f^a)$ induces a well defined operator on $\F_{\rm s}(\H)$.
Clearly, if the pair $(g_a,f^a)$ belongs to the 1-particle Hilbert space
$\H$ the answer is in the affirmative. We shall take this criteria also as
necessary condition. The question is now whether the pair  $(g_a,f^a)$
defines an element of $\G$, namely, whether they are `well behaved'
initial data for a solution of Maxwell equations with finite norm. This
will be the case iff the norm of  $(g_a,f^a)$, given by Eq. (\ref{norm3}),
is finite. This question is of relevance when defining  
observables given by
the fluxes of electric and magnetic field across surfaces bounded by
closed loops. The Heisenberg uncertainty principle takes a particular simple
form when this observables are considered \cite{ac}.


\subsubsection{Self-dual Decomposition}

As we mentioned in the last section, one can define the dual tensor to
the electro-magnetic field tensor $F_{ab}$, by
${}^*\!F_{ab}:=\frac{1}{2}\epsilon_{abcd}F^{cd}$. Note that if we
apply the duality $*-$operator again we get:
\ba
{}^*\!({}^*\!F_{ab})&=&{\T\frac{1}{4}}\epsilon_{abcd}
\epsilon^{cdef}F_{ef}\nl
& = & - F_{ab}\, ,
\ea
since $\epsilon_{abcd}\epsilon^{cdef}=-4\delta^e_{[c}\delta^f_{d]}$.
Therefore, the $*-$operator defines a complex structure $J$ on $\G$.
Note that this structure is available for any 4-dimensional Lorentzian
manifold $(M,g_{ab})$ without the need to introduce extra structure.
As discussed above, the $*-$operation decomposed the complexification
of $\G$ into eigenspaces with eigenvalues $\pm i$. The elements of
$F^{\uparrow}_{ab}$ of $\G_{\C}$ such that
${}^*\!F^{\uparrow}_{ab}=iF^{\uparrow}_{ab}$ are called {\it self-dual};
and those that satisfy ${}^*\!F^{\downarrow}_{ab}=-iF^{\downarrow}_{ab}$
are {\it anti-self-dual}. The corresponding projector is given by,
\begin{equation}
{K^{\uparrow}_{ab}}^{cd}={\T\frac{1}{2}}(\delta^c_{[a}\delta^d_{b]}-i
{\epsilon_{ab}}^{cd})\, .
\end{equation}
Therefore, the self-dual electro-magnetic field is of the form:
$F^{\uparrow}_{ab}=\frac{1}{2}(F_{ab}-i{}^*\!F_{ab})$. In terms of
objects defined on the hyper-surface $\Sigma$, namely electric and
magnetic fields, a self dual element is of the form $E_a-iB_a$.
Let us now write the projector $K^{\uparrow}$ acting on the pairs
$(A_a,E^a)$,
\begin{equation}
K^\uparrow\cdot \left(A_a\atop E^a\right)=\frac{1}{2}\left(A_a+id_a \atop
E^a -i B^a\right)\, ,
\end{equation}
where $d_a$ is the {\it electric vector potential}, i.e.,  such that
$E^a=\epsilon^{abd}\partial_bd_c$.

Finally, we could follow the same steps as in the previous case and
write the `norm' in the 1-particle Hilbert space
constructed from the $*-$operator decomposition as follows,
\begin{equation}
\langle(A,E),(A,E)\rangle=-\frac{1}{2\hbar}\int_\Sigma \d^3\! x\,
(E^ad_a+A^aB_a)\, .
\end{equation}

Note that this norm, in contrast to the positive frequency decomposition
case, is not {\it positive definite}, and is therefore, physically incorrect.
In math jargon, one says that the complex structure defined by the
$*$-operator is not compatible with the simplectic structure.
If one were to quantize naively this
``Hilbert space'', one would get a Fock representation with negative norm
states. In spite of this, it is possible to quantize the system when
dealing with self-dual fields.
 A holomorphic quantization with a positive definite inner
product was constructed in \cite{rov-sm}, and the corresponding loop
representation is the subject of \cite{8}.

\section{Discussion}
\label{C.3}

In this paper, we have introduced the Fock quantization for the classical
Maxwell field. We have seen that given a phase space point, that is,
a solution to Maxwell equations on space-time (or equivalently, a pair
$(A,E)$ of initial data), we can construct a quantum state via a
creation operator. There are several questions that come to mind. First,
How can we make contact with the ordinary treatment of Fock spaces
given in textbooks? Recall that, from the outset, the basic
fields are written in a Fourier expansion. This already assumes a
vector space structure for the background space-time (Minkowski) and
a globally defined vector field (time coordinate) in order to perform the
Fourier transform. The expression (\ref{norm3}), when re-expressed in
the Fourier components takes the familiar form of the inner product
found everywhere. This proof is left as an exercise for the reader. 

Second, we can ask how is that the particle interpretation of the theory
arises? We have used solutions to Maxwell equations to create
the `n-particle states',
but a classical electro-magnetic field certainly does not look
like a particle. Let us recall how it is done in ordinary textbooks. In
that case, the solution to the Maxwell equations is written in terms of
a plane wave expansion (via a Fourier transform), and each plane wave
with wave vector $\vec{k}$ is interpreted as (the wave function) of a
photon of momentum in the $\vec{k}$ direction. Thus, the Fock space is
constructed from plane waves, each with the interpretation of a
`particle'. Strictly speaking, plane waves are not normalizable and,
therefore, do not belong to our phase space $\Gamma$. 

Finally, we can ask  
how  the Fock quantization compares with the standard
Schr\"odinger representation we are used to in ordinary quantum mechanics.
Recall that in this case, quantum states are given by complex-valued
functions on configuration space $\psi(q^i)$. There is however, a
unitarily equivalent representation where the wave functions are
(analytic) functions on phase space $\phi(z^j=q^j-ip_j)$. This is the
so called {\it Bargmann representation} of quantum mechanics. This is
not usually done in ordinary quantum mechanics,
but we could in fact construct a Fock space for the harmonic oscillator,
where the `particles' would be quanta of energy  \cite{wald2}. In this
case the basis is given by the $|n\rangle$ kets, corresponding to the
eigenstates of the Hamiltonian.
The most natural
representation for this construction, in terms of wave-functions
 is the one given by Bargmann.  
Thus, the Fock representation is the field theory analog of the
complex Bargmann representation (for details see \cite{aa:am80}).
Is there in field theory the analog
of the Schr\"odinger representation? Can we construct it?
The answer to both questions is in the affirmative.
In the Schr\"odinger representation, quantum states
are functionals of the potential $A_a$ on $\Sigma$, $\Psi(A)$ and the basic
observables (\ref{observ2}) and (\ref{observ3}) are represented
as derivative and multiplicative operator respectively \cite{kuchar}.
Just as in ordinary
quantum mechanics, where the Schr\"odinger and Bargmann representations are
connected by a {\it coherent state} transform, there is a similar
transformation in field theory relating Schr\"odinger and Fock states.
Which of this representations is more useful? The answer depends on the
situation. Fock representations are very useful when considering
scattering processes. In perturbation theory one considers incoming
free states and outgoing free states (belonging to the Fock space) and
one tries to approximate the Scattering matrix relating them using a
perturbative expansion.
The problem with this approach,
from the mathematical viewpoint, is that this
procedure is not completely justified \cite{haag}. To explain
why, then, perturbation theory is so succesful is still an open problem.
The natural way to construct a quantum theory for non-linear fields is
then the Schr\"odinger representation (or its path integral variant), but
progress in this direction has been slow \cite{glimm}.

\section*{Acknowledgments}

The author would like to thank A. Ashtekar for discussions, the referee
for helpful comments and DGAPA, UNAM for financial support.


\begin{thebibliography}{99}

\bibitem{textbook} See for example: M. Kaku, {\em Quantum Field Theory}
(Oxford University Press, Oxford, 1992); J.D. Bjorken, S.D. Drell, {\em
Relativistic Quantum Fields} (McGraw-Hill, New York, 1964); F. Mandl, G.
Shaw, {\em Quantum Field Theory} (John Wiley \& Sons, New York, 1984);
C. Itzykson, J.B. Zuber, {\em Quantum Field Theory} (McGraw-Hill, New York,
1977).

\bibitem{tate} A. Ashtekar, R.S. Tate, {\em J. Math. Phys.} {\bf 36},
6434 (1994).

\bibitem{wald2} R.M. Wald, {\em Quantum Field Theory in Curved Spacetime
and Black Hole Thermodynamics} (Chicago University Press, Chicago, 1994).

\bibitem{wald1} R.M. Wald, {\em General Relativity} (Chicago University
Press, Chicago, 1984).

\bibitem{penr} R. Penrose, in {\em Battele Recontres}, ed C. DeWitt and
J.A. Wheeler, (Benjamin, New York, 1968).

\bibitem{penr2} R. Penrose and W. Rindler, {\em Spinors and Space-time}
(Cambridge University Press, Cambridge, 1984).

\bibitem{geroch2} R. Geroch, {\em Mathematical Physics} (Chicago University
Press, Chicago, 1985).

\bibitem{covariant} C. Crnkovic and E. Witten, in {\it Three houndred
years of gravitation}, Cambridge U. Press (1987); A. Ashtekar, L. Bombelli
and O. Reula, in {\it Mechanics, Analysis and Geometry: 200
Years after Lagrange}, Francaviglia Ed., Elsevier Science Publisher (1991).

\bibitem{dirac} P.A.M. Dirac, {\em Lectures on Quantum Mechanics} (Yeshiva,
New York, 1964); M. Henneaux and C. Teitelboim, {\em Quantization of Gauge
Systems} (Princeton U. Press, 1992).



\bibitem{6} A. Ashtekar and A. Magnon,  {\em Proc. R. Soc.}
(London) {\bf A46}, 375 (1975).

\bibitem{7} B. S. Kay, {\em Commun. Math. Phys.} {\bf 62}, 55 (1978)

\bibitem{aa:am80} A. Ashtekar and A. Magnon-Ashtekar, {\em Pramana}
{\bf 15}, 107 (1980).

\bibitem{geroch} R. Geroch, ``Special Problems in Particle Physics'',
 (unpublished).

\bibitem{reeds} M. Reed and B. Simon, {\em Functional Analysis} (Academic
Press, London, 1980).

\bibitem{ac} A. Ashtekar  and A. Corichi, {\em Phys. Rev.} {\bf D56}, 2073
(1997).

\bibitem{rov-sm} A. Ashtekar, C. Rovelli and L. Smolin, {\em J. Geom. Phys.}
{\bf 8}, 7 (1992).

\bibitem{8} A. Ashtekar and A. Corichi, {\em Class. Quantum Grav.}
{\bf 14}, A43 (1997).

\bibitem{kuchar} K. Kuchar, {\em J. Math. Phys.} {\bf 11}, 3322 (1970).

\bibitem{haag} R, Haag, {\it Local Quantum Physics, Fields, Particles,
Algebras} (Springer Verlag, Berlin, 1992); R.F. Streater, A.S. Wightman,
{\it PCT, spin statistics, and all that} (W.A. Benjamin, New York, 1964).

\bibitem{glimm} J. Glimm and A. Jaffe, {\it Quantum Physics, a Functional
Integral Point of View} (Springer Verlag, Berlin, 1987).

\end{thebibliography}
\end{document}